\documentclass[12pt]{article}

\usepackage{graphicx}
\graphicspath{{figures/}}

\usepackage{amssymb}
\usepackage{amsmath} 

\usepackage{algorithmic}
\usepackage{array}
\usepackage[caption=false,font=footnotesize]{subfig}
\usepackage{lipsum}
\usepackage{hyperref}

\begin{document}

\title{Conversations for Vision: Remote Sighted Assistants Helping People with Visual Impairments} 
\author{Sooyeon Lee\textsuperscript{1}, Madison Reddie\textsuperscript{1}, Krish Gurdasani\textsuperscript{2}, \\
Xiying Wang\textsuperscript{1}, Jordan Beck\textsuperscript{1}, Mary Beth Rosson\textsuperscript{1}, John M. Carroll\textsuperscript{1} \\
School of Information Science and Technology, the Penn State University\textsuperscript{1}\\
Department of Computer Science, the Penn State University\textsuperscript{2}\\
}

\date{}

\maketitle 

\begin{abstract}
People with visual impairment (PVI) must interact with a world they cannot see. Remote sighted assistance has emerged as a conversational/social support system. We interviewed participants who either provide or receive assistance via a conversational/social prosthetic called Aira (https://aira.io/). We identified four types of support provided: scene description, performance, social interaction, and navigation. We found that conversational style is context-dependent. Sighted assistants make intentional efforts to elicit PVI's personal knowledge and leverage it in the guidance they provide. PVI used non-verbal behaviors (e.g. hand gestures) as a parallel communication channel to provide feedback or guidance to sighted assistants. We also discuss implications for design.
\end{abstract}

\section{Introduction}

Recent years have seen the rise of remote sighted assistance for people with visual impairment (PVI). The world is not always accommodating of visual impairment, and PVI often need help. Many PVI worry about burdening their loved ones, and strangers are not always helpful. Remote assistants give PVI the opportunity to utilize their own abilities while receiving supplemental information from a trained individual whose job it is to provide support to PVI. 

One notable example is a company called Aira~\cite{aira}, which is an on-demand subscription service for remote sighted assistance via video chat. There are services similar to Aira, like BeMyEyes~\cite{bemyeyes}, TapTapSee~\cite{taptapsee}, and VizWiz~\cite{bigham2010vizwiz}, but we have found Aira to be more robust and preferred by PVI that have used it as well as other companies. Aira's trained human agents receive calls from PVI and proceed assist them with a wide range of activities through conversational interaction. Agents have access to the video feed from PVIs' phone camera or smart glasses, their GPS location, and Google maps. With these tools, they process and evaluate information about the PVI's environment and present them with information deemed relevant based on urgency, goal, and PVI preferences. PVI can put the agent on speaker or be more discreet by using headphones or Bluetooth. 

Aira is a highly accessible technology that uniquely keeps PVI in control of the collaboration while still connecting them with a human assistant. For these reasons, it is an extremely interesting, rare, and valuable subject of study. For this research, we developed a relationship with Aira and were granted access to their agent employees and their PVI users for interviews. By studying Aira, its success, and its use of conversational support for PVI from the perspectives of both parties involved, we seek to answer the following research questions: (1) What kind of communication/interaction is needed for remote human agents to assist PVI? (2) How should information presentation by agents differ from context to context/task to task? (3) What are the implications of our findings for the design of conversational user interfaces for PVI and PVI-sighted collaborative technologies? 


\section{Related Work}
\subsection{Photo/Video-Mediated Interaction}
Investigation into the effectiveness and feasibility of crowdsourced human assistance for helping PVI to understand their surroundings is an interesting field of research. Research prototypes and commercial products have allowed PVI to take pictures which either AI or a remote human then describe. VizWiz is one well-recognized example in assistive technology research that utilizes this method. They have also used their initial findings to further develop and expand their work through VizWiz Social, which connected even more PVI to remote assistants through social networks. There are drawbacks of VizWiz, though, with many of PVI's questions being deemed unanswerable by remote workers~\cite{gurari2018vizwiz}. Researchers found that PVI actually want to partake in a more interactive collaboration with agents, and the results of VizWiz studies have demonstrated the value of a live video feed and a back-and-forth conversational interaction~\cite{bigham2010vizwiz}.

BeMyEyes is another similar service. However, the extent of the interaction between assistants and PVI and the scope of help they are able to provide are very limited. BeMyEyes uses crowdsourced volunteers, which are not always familiar with the needs of PVI and are not trained. PVI cite this as a problem. 

Some studies have shown PVI using their personal social networking platforms to ask questions about pictures they have taken. This method usually has a low success rate and is associated with social costs for PVI. Many PVI are reluctant to ask for help on social media because they do not want to burden their peers and they worry about revealing sensitive information~\cite{brady2013investigating}. 

\subsection{Sighted-PVI Collaboration}

Conversational interactions between PVI and sighted assistance involve socializing, clarification, and confirmation. Important characteristics of these working relationships are PVI's trust in the sighted individual, mutual contribution, and agreement on how the task at hand should be executed. Continuous communication and clarification keep these well-established throughout the interaction~\cite{carroll2003notification}. Formation of common ground at the beginning of the interaction also paves the way for a successful and respectful collaboration. Three sources of common group have been identified: assistance knowledge, interpersonal knowledge, and domain knowledge~\cite{yuan2018idid}.

These are especially important when the two parties involved do not have access to the same information, like with remote sighted assistance of PVI. As a remote agent, the assistant does not have the same degree of immersion in the environment as PVI, but this obstacle can be overcome by the foundations described above~\cite{yuan2018idid}.

\subsection{Conversational assistance for PVI}

Conversational interaction with PVI should not be approached the same way as sighted-sighted conversation. PVI rely almost exclusively on auditory input to process their environment, so they prefer to receive much more speech than someone who is sighted and has other methods of gathering information would~\cite{coroama2003chatty}.
This difference should always be a primary consideration in conversational interaction design

Anam et al.~\cite{anam2014expression} has used Google Glass to transmit nonverbal expressions to PVI during their conversational interactions with present 3rd parties using AI. 

Tanveer et al.~\cite{tanveer2013you} also investigated the use a visual-to-auditory Sensory Substitution Device for helping PVI understand facial expressions. The goal of this is to enhance PVI’s social awareness and communication skills.

\section{Method}

\subsection{Participants}
We have established a partnership with the company Aira, which provides PVI with assistance from wearable technology and a trained, remote human agent. We recruited both agents and users for interviews with the help of Aira. A recruitment letter and an online recruiting form created by the research team for potential participants was distributed to the agents and users by the company. We recruited a total of 12 study participants, including four agents and eight users. The agents are comprised of one male and three females, and their ages range from 23 years to 50 years old. They all had some degree of experience assisting Aira users. The user participants included two males and six females, and their ages range from 20 years to 65 years old. All of the Aira user participants are currently students. Please see the participant information tables~\ref{tab:ParticipantDemographics,tab:ParticipantDemographics2} for further demographic and personal details. 

\begin{table*}
  \caption{PVI Participants}
  \label{tab:ParticipantDemographics}
  \begin{tabular}{p{1.2cm}cclp{4.5cm}p{3.5cm}p{1cm}}
    \hline
     - & Gender & Age & Age of Onset & Condition of vision impairment & Occupation\\
    \hline
    Kate & F & 32 & At birth & Premature, congenital & Student (political science major) \\
    \hline
    Cole & M & 32 & 4yrs & Legally blind, 20/700, shaky eyes & Student (Computer science major) /consulting for family and friends, beta testing app\\
    \hline
    Robert & M & 62 & 39yrs & Totally blindness & Student/peer support specialist library social worker (scholarship plan)\\
    \hline 
    Linda & F & 23 & At Birth & Usable vision on right eye -   color, shade & Student & \\
    \hline
    Irish & F & 19 & At Birth & total blindness,
a little bit of light perception & Student (special education/ vision major - dual major) \\
	\hline
	Jennifer & F & 25 & At birth & Total blindness & Student (family studies)\\
    \hline
    Heidi & F & 35 & At birth & Total blindness, ROB & Student/technical advisor \\
    \hline
    Shelly & F & 27 & At birth & ROB and Glaucoma; light perception on both eyes& Student/technology instructor\\
    \hline
  \end{tabular}
\end{table*}

\begin{table*}
  \caption{Agent Participants}
  \label{tab:ParticipantDemographics2}
  \begin{tabular}{p{1.2cm}ccp{3.5cm}p{7.5cm}}
    \hline
     ID & Gender & Age & Time as an agent & Previous experience with PVI \\
     Stella & F & 25 & 10 months & None \\
     Chris & M & 50 & 1.5 months & Helped people worked to regain lost vision in the past\\
     Ginger & F & 22 & 10 months & Some volunteer work \\
     Kimberly & F & 24 & 1 year & None \\
     \hline
  \end{tabular}
\end{table*}

\subsection{Procedure}
We conducted all agent and user interviews over the phone. We read a consent form to each interviewee at the beginning of the interview and obtained verbal consent to proceed. The interviews lasted from 45 minutes to one hour and were audio-recorded in their entirety. We used open-ended, semi-structured questions as a guide for conducting the interviews with both the agents and the Aira users. We asked the agents to share their experiences assisting PVI remotely, specifically with regards to strategy for relaying information; managing and overcoming challenges; methods of communication with PVI; and finally, how the aforementioned vary for different tasks, situations, and contexts. Similarly, we asked users about their experiences utilizing the Aira assistant, focusing on the interaction and communication with agents and how these change from case to case. While users described their experiences, we paid special attention to what they like about the service. We also added one interesting question regarding how they choose which assistive technology to use for different tasks. In addition to the audio records, the interviewer made notes during the interviews to capture as much information as possible.

\subsection{Data Analysis}
All twelve audio-recorded interviews (four agent and eight PVI interviews) have been transcribed.
The data was analyzed with a bottom-up, iterative, thematic approach~\cite{braun2006using}.
The interviewer conducted the first round of coding analysis on the agent interview data.
Based on the initial coding, using the interview questions as an anchor, high-level themes emerged and were grouped into more general categories. The necessary kinds of assistance, modes of interaction and communication, strategies and knowledge to be used, challenges, and the fact that all of these factors ought to vary based on the situation at hand are all common themes among the Aira agents. Another researcher performed a parallel coding analysis on the same data and additional themes uncovered were incorporated into the findings.
The same iterative coding and thematic analysis method was used on the user interview data. After extracting the recurring themes, the research team discussed the results and constructed a set of requirements for conversational agents for PVI. 

\section{Findings}


\subsection{Conversation in Scene Description} 
\subsubsection{Task/Context} 
We found that the object identification and description-related contexts that PVI sought assistance with the most include the following tasks: reading slides/text; describing spaces, scenery, images, and people; and identifying objects. These tasks require the agent to present visual information to PVI in a way that is conducive to the individual's formation of a mental representation of the scene.

\subsubsection{Agent requirements}
In providing PVI with a description of an object or general surroundings, the agent needs to determine the best mode of information delivery in terms of vocabulary and level of detail. We learned from our interviewees that the agent needs knowledge about the following: (1) the severity of PVI's individual visual impairment, (2) the onset of the visual impairment, (3) what type of experience PVI have, and (4) topical information like science or math, if applicable. We found degree of detail to usually be determined by the task type and the type of information requested, but this is also negotiated and fine-tuned during the interaction. 

\subsubsection*{Analogies and terminology for PVI}


We found that knowing the general time of onset of the PVI's visual impairment is very helpful in the agent's decision about how to communicate with the PVI. This information is meaningful since the knowledge and experience that the PVI had before losing his or her sight form the basis of their understanding of the visual world, and agents and PVI can both use this as a reference point. For example, providing color-related information to someone who has been blind since birth is completely arbitrary. One agent interviewee, Stella, said,

\textit{''One of the things we ask them for in the signup is how long they have been blind.  That helps us.  We might ask them if they have been blind 5 or less years or more than ten years. If I see that someone has been blind for five years or ten years, I know I can talk about color or talk about different nuances. If I meet someone that is blind their whole life, I can use different descriptions like shape. It's not simplistic, but maybe focus more on the shape rather than spending so much time to talk about color because for them it is not as applicable.''}


She also added that having knowledge of specific topics is very helpful and implied that this is one of the challenges she faces in her work. 

\textit{''Sometimes there are things like... it's more specific. For example, someone is in like a neuro class like something more microbiology or neurophysiology, things like that.  I have had a couple people call in for specific diagrams like 'the teacher sent us this, we need to know it, but what am I looking at?' (PVI).  Verbally, I describe everything I can, for example like the cross-section of a brain, and describe everything that I see there...I thought I wish I could have either had more experience in that area or be able to describe the material more clearly because it does take her a while to understand (Inaudible) or where the inhibitor was.  And so I think that was a little bit challenging, so I felt like I could have done a little bit better because I wasn't quite satisfied there... describing a specific diagram, I don't always quite feel satisfied afterwards.''}


One Interviewee, Ginger, shared her story of assisting  PVI using analogies. We were impressed with how she discovered something she had in common with the PVI during their conversation and how she utilized it in an analogy.

\textit{''When someone was taking, I think an advanced genetics class, that they were getting into the gene and the way it works and it was very complex. There was a lot of aspects to it that she had no concept of what it looked like so I was trying to describe it to her and the only thing I could think of was each part and comparing it to pasta noodles and well this {Inaudible} looked like a macaroni noodle and this one is an angel hair spaghetti. That's what they look like and comparing it to that helped a little bit. I think I mentioned something about sewing and this one looks like a sewing needle and she said 'oh I love sewing' (PVI), and after that I used sewing analogies like 'this one looks like a pin cushion' (Agent G) and 'this one looks like a thimble' (Agent G) just to try to make it a little bit more familiar to her.''}

\subsubsection*{Detail level of descriptions} 

The level of detail used in descriptions seems to be determined by what purpose the description serves to the PVI. The agents stressed that it changes by task, situation, and individual PVI. We found that it is adjusted on the fly as the interaction unfolds. The agent uses the following to inform their adjustments of the detail level of the information: (1) direct requests from the PVI, (2) confirmations by the agent, and (3) non-verbal expressions from the PVI.

One agent participant shared his experience with us, saying that one particular PVI he assisted told him that he did not need to know that much detail and asked to be given less detailed information. 

\textit{''So this person's profile said that they like a lot of description and so he was walking down the street and I was giving a lot of description and he actually stopped me and said, 'Stop giving me so much detail. You are giving me too much detail.' Even though his thing was saying give him a lot of detail, there was a bit of interchange between us to understand what that really meant for him which is, 'I need some detail, but you are giving me so much detail I'm confused, and I don't understand what you are saying.'''}


When a PVI participant, John, was asked about how to adjust the detail level of descriptions the agents give, he said he too simply lets the agent know if it is excessive. 

\textit{''That is enough buddy, let's keep going.''}

\subsection {Conversation in Navigation}
The most common task requiring directional guidance identified was navigation. Navigation entails both directions and description of the environment. PVI frequently ask for a navigational assistance when trying to find a gate in an airport, navigating on campus, and finding a location in an unfamiliar area. These situations are different from other contexts in that there are many unpredictable variables that are constantly changing, so fast, real-time information delivery is needed. Stemming from the characteristics of this context, the mutual goal of navigation in space for both the agent and PVI is to get to the desired location safely and efficiently. 

\subsubsection{Challenges in temporal coordination}
Due to the marriage of directional instructions and scene description involved in navigation, we found that the biggest challenge for all of our agent interviewees, especially on public roads, is having to deliver too much information in real-time as well as keep up with the changing environment and movement of the PVI. 

One agent, Chris, expressed his nervousness and stress associated with this specific task and said that some PVI move faster than he can describe the environment: 

\textit{``They are cranking, they're walking faster than people who are sighted, I actually need them to slow down. Words take time to express meaning... when I'm walking down the street, I don't need to say to myself, `there's a trash can in front of you, it's rectangular, about waist high.'}''

Another agent, Ginger, also described the difficulty of such a situation:

\textit{``It can be overwhelming, when everything is moving quickly, when a PVI is walking really fast, it can be hard to see everything that is approaching or passing by.''}

A third agent, Kimberly, told us about how she can adjust to PVIs' rapid paces: 

\textit{``If I notice that someone is walking really fast, I will try to describe things a lot further ahead of them, so that I can make up for that latency, so that I am not too far behind. We have some travelers that are just like lightning and they move so fast that I'm just trying to just keep up. You just try to describe it earlier/sooner and talk a little bit faster.''}


To cope with this challenge, agents can change their pace and prioritize certain information. If the PVI is walking down a crowded street, obstacle-related information will be the agent's focus. If the agent senses urgency, they will focus on providing PVI with directional instructions to help them get to the destination as fast and safe as possible and skip some of the unnecessary scene description. Stella said: 

\textit{``I immediately try and figure out what is the level of emergence in this call. How fast do they need me to help them?''} 

Throughout the conversation, the agent and the PVI collaborate to adjust the PVI's speed,  the level of detail in descriptions, and the urgency of the task through verbal and nonverbal communication.  




\subsubsection{Silence Rule: Being silent while PVI use orientation and mobility skills} 

Interestingly, we found that agent's silence is a deliberate piece of the interaction between the agent and the PVI. When PVI are crossing an intersection, the agents are not allowed to speak and or provide any assistance. The rationale behind this rule is avoiding interfering with PVI's orientation and mobility (O and M) training. PVI are trained to rely on their auditory sense to avoid potential danger (e.g. listening for cars crossing an intersection). An agent's speech can divert some of the PVI's attention or overload their auditory sense and cause them to misjudge dangerous situations. Therefore, Aira's protocol is that agents must refrain from speaking while PVI cross the street. It also respects their independence and capabilities. 

\textit{``All of these things that they have been taught, to navigate crosswalks and things like that, that is connected to independence. I can be in the world myself, and Aira is a partner in that , but we never want to take that away from them, because then it takes away that independence. And sure, maybe it’s easier, there might be an explorer who likes being just told do this, this and this but then it takes away their independence and actually puts them more in danger	especially if you are crossing the street.''}

To one agent, Chris, it is one of the stressful and difficult situations. Remaining silent from start to finish even if he sees PVI deviating off course or stuck in the middle of an intersection is hard to do. He described how this has become a problem for him when guide dogs make mistakes and drag PVI into traffic.

\textit{``Sometimes even if they are relatively lined up right, they wander, they go off course. Twice I have had a guide dog literally pull the person into the middle of the intersection, because the guide dog got confused, and I can't say anything. It's like, they are going to have to figure this out and there are literally cars that are screeching by them and the dog is trying his best, it just got a little confused.''}

\subsubsection{Trusting the agent's instruction and being confident in their skills} 
We found that conversations centered on navigation are a bidirectional information channel. In this context, we observed that leading and following roles emerge. In a leading and following relationship like this, we found that trust in the partner who provides instruction plays a very important role in making the conversation successful. Stella told us how crucial PVI's trust in her is for her work: 

``\textit{So for navigating, you can tell they're confident when you tell them an instruction and they're like ``Okay'' and they just start moving. Some (PVI) are very timid and they're scared they're going to run into things. They're scared  they're going to step off the curb and into the street... I think confidence in their skill and confidence in knowing they have a dog or a cane to help them.  Also, just confidence that `You know what, hey I might fall off the curb and into the street but that's okay, it might happen.'}''

\textit{``Oh I love when someone is confident! It makes my job so much easier. They're more responsive. I was helping someone in a... She started walking and I said turn a little bit to the left.  Sometimes when you tell a (PVI) that, they'll stop slowly and turn and they'll keep walking and say `am I going the right way'(PVI) where as people who are confident will turn left. Then if you say `turn a little bit to the right,' they'll turn right and just go. They're just so responsive and confident in their skills and trusting in me that I'm not going to let them run into anything.  It's just really nice. I love it.''}

\subsubsection{Knowing the other's level of familiarity with the space.}
We found that familiarity with and knowledge of the location and area of both parties is another important factor in their coordination. Agents use this in their decision of what type of information to present and how much detail to include. PVI and agents mutually share their knowledge to help one another; it is not just the agent giving information to the PVI. One PVI, Shelly, talked about how she shares what she knows about her college campus with agents who help her navigate to make their job easier:
\textit{``The more I know about the campus, the more I can help the next agent.''}

It applies equally on the PVI side. The PVI's understanding of how much the agent knows about the area they are in will help make interaction more productive and pleasant. Even though the agents have maps on their screen and the PVI's location, they have no access to the real environment where things are constantly changing in 360 degrees, and they can only see the PVI's camera feed. The agent interviewees spoke about having a hard time with such limited access to complex surroundings. The agent Ginger said: 

\textit{``Airports are just designed willy-nilly and maps don't make sense, it's hard to navigate when you're not there''.} 

Another agent, Kimberly, expressed how difficult navigation can be for the agent, highlighting an example of assisting PVI in a city:  

\textit{``'Okay, I am outside my apartment and I want to walk over to the seven-eleven, that I know is about three blocks away,' (PVI). So, you begin to help them navigate that and just the... You want to shake the city planner... it’s lovely they put all these potted plants and different things and the sidewalk and sure, it's lovely in design and a nightmare to navigate, because for a sighted person, sure I can walk by and around all those things that are in my way. But if I didn't have sight, or even, regardless of sight, if I had a mobility issue, that might be hard to navigate around so many obstacles, even though it looks pretty in the sense of environmental design.''} 

Chris also shared his experience with this and expressed his frustration stemming from PVIs' expectations that he be familiar with their environment and be able to immediately start providing assistance. Consequently, the interaction was marked by tension and conflict. He said he needed some time to orient himself and familiarize himself with the new environment, and he said he needed to negotiate with the PVI regarding this matter:

\textit{``When he (PVI) calls in, he is assuming that the person that is helping his can immediately just translate everything as if they have been there all their lives. I am walking down the street, and if I am helping him, I may never have walked down that street. I am not only describing things but I am having to process information that I am seeing through goggles.''}

\subsection {Conversation During Performance}
We found that PVI also ask for agents' assistance for performing tasks such as doing a live presentation; teaching a class; or making presentation slides, banners, and resumes. Compared to the previous two situations (scene description and navigation), the interaction between the agent and the PVI here is more intimate, personal, and dynamic. This task involves scene description, instructions for action, as well as more detailed coordination. How much information, when to provide it, and how it is to be communicated needs to be agreed upon by the agent and PVI. 

\subsubsection{Communication protocol setting and practice: Live presentation} 
It was interesting to find that the agent and the PVI set up a communication protocol and rehearse before actual presentations. An agent, Stella, detailed the process of how she helped a PVI prepare: 

\textit{``Beforehand, I went through each of the slides with her and then we kind of talked about them.  So, she said `these are the things I need to know about the presentation' or `these are the key words or the key facts I want to know.'  So then when the presentation started, I just kind of gave her words she needed. For example-it was about traffic, she wanted to talk about traffic. Right as a slide came on I said `traffic and two pictures' and so she knew there were two pictures that came on and she knew to talk about it. Or `This one is about popsicles' And then from there she knows this is the slide to talk about popsicles.''}

\subsubsection{Discreet conversational interaction: Live presentation in front of audience}
In doing live presentation in front of people, we discovered that the PVI want to demonstrate that they are able to carry out a presentation by themselves without help. To do this, they develop unique means of communication that the agent then learns, and they are able to confidentially converse during the presentation (e.g., forgetting what on the slide). One agent, Stella, said: 

\textit{``That is something they (PVI) have developed and that is just communication I have learned to work with. Sometimes I could tell she was struggling to figure out what's on her slide, but because she had earbuds on or Bluetooth, no one else could hear what I was saying to her. It was completely private.''} 

\subsubsection*{Various communication methods and unique usage}
The unique communication means that we found to be used include pausing and cueing. During a presentation, explicit verbal communication is not an option for PVI. Therefore, they use other modes like pausing and cueing, which alerts the agent that they would like to be fed information. The agent Stella supports this finding by saying: 

\textit{``I found a lot of really well-spoken blind people have learned to develop it [the pause]. They'll pause or take time to say a specific word.''} 

We found the ways that PVI use these non-explicit cues very interesting. Often times, PVI employ them when they forget or miss something. Stella continued: 

\textit{"They're pretty good at knowing when they have missed something.  So then they'll pause or cue me like `And the second point on this iiiiiiiiiiiiiiiiiiiiiissssssss' and then they'll pause and I will say `popsicle' as quickly as I can. Then they will continue and say `its popsicles!'"}

\subsubsection*{Feeding information during pauses}
There are limitations associated with live presentations and using discrete communication methods. Agents must pay close attention, wait for cues, and react to them in timely manner. They also refrain from interrupting the presenter even if they have missed something. Rather, they know to wait for the PVI to pause so as not to throw them off. Agents have found that following these protocols promote a successful presentation: 

\textit{"Typically I will wait for a pause, I won't interrupt (PVI). If they move on to the next slide I will say `popsicles but on the last slide.' They will say `oh by the way I forgot' and they will go back and talk about it.  They won't switch slides they will just follow up on it.  And It seems normal because even sighted people do that. They will say oh yeah, 'and I forgot to mention this blah blah blah.' I don't interrupt ever, I find that very distracting."} 

PVI also have to be strategic in listening to the agent while not making it obvious to their audience: 

\textit{"They'll pause or take time to say a specific word. They'll draw things out to make it more dramatic.  It sounds like it's totally part of what they're trying to do, but rather they're really just listening to me."}

\subsubsection{Discreet conversational interaction: Engaging with the audience}
In addition to assisting the PVI with presenting information, the agent also helps the PVI to engage with the audience. The agent gives description in such a way that PVI can interact with the audience as sighted presenter would, like call on audience members raising their hands. We found that pausing and asking questions can serve as a cue for when PVI want information from the agent.

\textit{"If she's asking a question like `Can anyone tell me what the common name of Tylenol is?'(PVI). I'll wait and say if someone is raising their hand. I'll say `the gentleman on the left is raising his hand' and she'll say `you sir' (PVI). I will try to give a at least a little bit of a description about someone who is raising their hand and if they're wondering where they are in the room,  I'll tell them the direction, and a descriptor like if they're a man or woman, if they're wearing a red shirt, and they're sitting in the front. She'll say `you sir in the red' (PVI) and sometimes it's a joking thing because they know she is blind... or maybe they don't."} 

We also noted these descriptors need not be detailed. The agent Kimberly said: 

\textit{"Usually I try to make it as brief as possible."}  

Agents also indicated that figuring out what information to provide and when to deliver it is an important and challenging task where formal interaction between the PVI and 3rd parties is occurring. An agent said: 

\textit{"I will think through beforehand what I will say because there are pauses where I'm not saying anything. I don't have much time.  People pause to talk (Inaudible). I think what can I tell this person in one or two seconds that will help them. So I think about it.  For me its just mental and planning..."}

\subsection {Remote Assistance in Social Contexts}

In some situations in which PVI request sighted assistance, there are 3rd parties present. This greatly affects the nature of the conversation between the agent and the PVI since they are no longer the only two people involved in the interaction. Sometimes strangers approach PVI while they are using Aira and offer help, and other times, PVI use Aira while with friends or family to avoid asking for their help and exercise their independence. In a social context, the information the agent presents and how and when they present it require some fine-tuning.  

\subsubsection{Three-party interactions}

In several cases, PVI have company while using Aira, and this 3rd party becomes an active contributor to the conversation. This can complicate things, especially on the agent's side. They then have to consider their effect on the 3rd party and the PVI's social experience and find a new equilibrium. One the one hand, the PVI has asked them for assistance, but on the other, the agent does not want to intrude in or hinder PVI's interpersonal interactions. Their primary concern is not speaking over the 3rd party. Stella, an agent interviewee, said: 

\textit{''Sometimes when I talk I'm being broadcasted, maybe not super loud, but I have to be more mindful of talking less. So that is a challenge, speaking as little as possible while also giving them effective experience essentially of helping them succeed in whatever they're doing and also feel satisfied with the service. I think the challenge is more social because there are other people around and it changes things. Maybe they want me to talk less.''}

Another agent, Ginger, emphasized the importance of \textit{''being respectful to the person who's there by allowing them to have their normal conversations.''}

In addition to adjusting how much they speak, we found that agents also seek to boost PVI's social awareness where they can. Stella continued: 

\textit{''Maybe there's people around so they want to know what the other people around them are doing or how they're feeling. I want to say things in a way that will help them look good in the eyes of their peers and also help them...socially, to do well.''}

Many of the visual parts of social interactions that can be taken for granted are actually vital to social success, and the same agent elaborated on how she realizes this and empathizes with PVI:

\textit{''Sometimes... they want to know the reaction of others in the conference room or pledge hall because that is important. If someone is proposing something, part of how we as humans make decisions is we'll see how is everyone else feeling about it. Is everyone excited about it or is this the worst idea ever? You don't want to be the only one. That's just natural peer pressure that humans feel. Part of it is that they (PVI) want to know how other people in the room are reacting to it as well.''}

In these cases, agents will spend more time describing things like the facial expressions and body language of the relevant people around. These can be complex and have nuances to them, so agents need to use a rich and descriptive vocabulary to paint a meaningful picture for PVI. One agent, Ginger, gave us some examples of the kind of descriptors she uses:

\textit{''Usually we could just say, like, if the person wasn't really making any facial expressions, it could be a neutral or resting face. If they were smiling, we would say that they were smiling, and it could be like, slightly smiling, or a toothy smile if it was a really big smile. Um, if maybe they were like looking up and kind of moving their head side to side, it would be kind of like a thoughtful expression, so just really describing what we see.''}

\subsubsection{Classroom settings}

A slightly more advanced social situation in which PVI have used Aira is in the classroom. Here, many of the above protocols still stand, especially avoiding speaking over others, as agents do not want to distract students from class material. There are also social implications for PVI seen receiving sighted assistance, which we found that agents are observant of. Stella said:

\textit{''It's different when someone in the classroom sees me talking... I feel like people tend to belittle visually impaired individuals... They treat them (PVI) like they're children in the way that they touch them, in the way that they address them and talk to them... They want to come across as professional and knowledgeable and adults, because they are adults. I think just being conscious of what I say, especially in a classroom with other people around. I think I mentioned this before, but just using my words to a minimum to where I'm not distracting but helping.''}

One way PVI have managed to be discreet while using the service is by wearing headphones and using hand gestures. Aira has established universal hand signals to be used by PVI that nonverbally communicates what they need from the agent if they are ever in a position where they cannot or prefer not to speak. Examples given include business meetings, lectures, and live performances (ie. plays). Kimberly explained the gestures:

\textit{''We have about three or four (hand signals) that we use on all calls universally, that we tell our (PVI) about, so if they give us a fist in front of the camera when we answer the call, we know that they're not going to be speaking. If they move their index finger left to right, we know they want us to read. If they give us an 'OK' sign, they want us to just describe, and then they can make a 'C' shape and that means to stop describing.''}

This allows PVI to behave as anyone else would while privately receiving supplemental information from an agent. It simplifies and also expedites the process of PVI getting what they need out of the interaction. Kimberly gave us her thoughts and an example scenario: 

\textit{''It works well. We don't have a ton of calls that are silent, but every once in a while, you pick one up and you just get the fist right in front of the camera and you go 'oh, okay I know you're not going to be talking, I'll go ahead and start describing or start reading.''}

\section{Discussion}

Visually impaired Aira users indicated to us that they are able to successfully complete a wide range of tasks with the help of the remote, sighted agent, which is no small feat. To make this possible, the interaction between the two parties is complex and variable. 

Aira is one of a few pioneers providing remote, sighted assistance to PVI, and though new, this practice is anything but simple. The communication channel is bidirectional, and agents and PVI go back and forth translating their knowledge into brief expressions that someone with a completely different way of experiencing life can immediately understand and respond to. 

It is not a one-dimensional relationship where the agent tells the PVI what to do. PVI do not rely on the agent, but rather, agents respect PVI's orientation and mobility abilities and autonomy, and PVI appreciate agents' supplemental knowledge. Together, the two form a connection built on a unique kind of trust and collaborate to achieve goals. 

\subsection{The emergence of a new prosthetic practice}
Through our interviews with both agents and PVI, we were able to extract some basic knowledge/skills that agents need to be effective remote assistants. Firstly, the agent should know enough about the PVI's impairment to be able to make decisions about what kinds of analogies to use and what aspects of the environment they should describe. One aspect of Aira that agents and PVI have cited as contributing to its success is PVI profiles. These contain the users' preferences and information about their frequently-travelled locations and are readily available for agents to check when receiving a call. Having some relevant topical knowledge can also be useful for helping PVI with things like reading music or making sense of a diagram in a biology class. While any sighted person may be able to process the image, it is not as easy to put something that you do not understand into words. 

A lot is demanded of agents, and their job is very unpredictable. Their most crucial asset, therefore, is adaptability. Agents are constantly thrown into situations where they are asked to explain something they have never seen or touched or give directions to a place they have never been. The PVI they assist also all have different travelling speeds and preferences for information presentation. Agents determine an approach at the beginning of the call based on the task and context and then continuously make adjustments throughout the conversation based on verbal and nonverbal cues from PVI. PVI have used Aira silently during lectures, on speaker during group projects, and privately at restaurants with friends around. Calls like these require agents to strategize and avoid speaking over 3rd parties and distracting PVI from their outside interactions. When only able to speak during pauses or if a PVI is moving very quickly, agents make snap decisions about the priority of bits of information and keep track of highly dynamic atmospheres. 

Also notable is the agents' deep empathy for PVI. Several of the PVI emphasized how much they could tell that the agents genuinely enjoyed their jobs and cared about the users. This especially came across when an agent, Stella, talked about how she tries to advance PVI socially and make sure that she gives them all the information they need to be socially adept where their impairment might make that difficult. Chris also spoke about supporting PVI in another capacity, saying: 

\textit{''Part of our work... is bearing witness to what they experience and being able to be someone on the other end of the line that is supportive... Even if I do not share the disability, I do share your experience.''} 

The agents clearly feel compassion toward PVI as they guide them through tasks and feel as though they are a part of that experience. Chris believes part of his role is an agent is a vicarious experience of visual impairment and standing with PVI throughout their joint experience. The agents have a stake in the PVI's experience, making them invested in the quality of their work.

\subsection{Limitations of human agents}
Most of the commonly-raised concerns regarding the use of human agents have been accounted for by Aira's protocols and infrastructure. Some limitations include finite information processing and delivery speed, limited topical knowledge, and inconsistency between agents. To address the first of these, agents learn to prioritize information and present PVI with information related to their safety, then directions, and then ambient description. Aira agents also have an internal network through which they are able to communicate with one another, and if an agent is assisting a PVI requesting topical information, they can ask other agents if they may be better able to help and pass the call off to a better-equipped agent. Finally, there are individual stylistic differences between agents, but our interviewees said that they primarily base their communication off of PVI preference, which they can easily access through the profiles. While there are always limits to human abilities, Aira has done a good job of identifying room for error and correcting for it. This model is an effective design template for similar concepts.

\subsection{Design implications}
From the experience of agents and the feedback of PVI, we have derived some design recommendations for conversational remote sighted assistance for PVI. When PVI compared Aira to similar services like BeMyEyes, they appreciated Aira agents' access to maps and Google. This way, agents were able to anticipate their location and provide them with preemptive information and directions. With BeMyEyes, the remote assistant gathers all of their information from the PVI's video feed. 

It also beneficial to PVI to be assisted by trained agents that are familiar with O and M terms and PVI-friendly vocabulary. BeMyEyes uses crowdsourced volunteers, and, as described above, PVI build mental models of spaces using different descriptors than sighted people. Untrained sighted assistants may not be aware of the altered lens they should be using to make descriptions meaningful to someone without vision. Even when specifically asked, PVI mentioned very few failures by Aira agents. 

The previously discussed PVI profiles are also part of our design recommendation. This brief of PVIs' preferences allow remote assistants to adapt their communication styles before the interaction even begins, which can save both parties the time and stress of negotiating this while working toward the goal. 

Even with this primer of PVI preferences, during the interaction, things may get a little off track, and it's important for the agent and PVI to have very open communication. If PVI are moving too quickly, the video feed is unclear, or agents are not meeting PVI's needs, each party should feel comfortable giving the other feedback and making mutual adjustments. Therefore, free and open continuous communication should be an underlying assumption in the interaction design. 

One improvement we offer is the incorporation of further nonverbal parallel communication networks. PVI interviewees suggested a texting feature be added so that they can get messages to agents without speaking and beyond what is permitted by the few established hand signals. We also recommend the addition of one-directional haptic feedback. If remote agents could alert PVI of an obstacle or directional cue in the time it takes to transmit a vibration to a device on the PVI as opposed to having to describe it in words, their stress level would decrease and they could provide PVI with more scene description. In situations where PVI are travelling quickly or there is a lot going on, PVI may still wish to know details about their surroundings, but as is, agents are not able to accommodate this. Haptic feedback would add a faster channel for agents to communicate simple bits of information. 

Based on our interview data, these design guidelines will make for an effective and enjoyable PVI-sighted agent conversational interaction. 

\subsection{Future work}
Little research into this intriguing new area has been done, and we have much more to learn from conversations between remote sighted agents and PVI. Aira has created a very successful platform for sighted assistance for PVI, but it is still imperfect. The agents and well-trained and well-practiced, but they can still only get so much information across in short time periods. Investigations into the use of conversational interaction layered with one-directional haptic feedback can inform the further improvement of services like this for PVI. Whether PVI are able to interpret haptic signals while also listening to a remote assistant and taking in sensory information from their immediate environment and whether this would be preferable is a pressing research question. 

We were surprised to find that some agents make it a point to help PVI be more socially aware. We learned that sighted assistants have the ability to provide PVI with opportunities to engage in and read social contexts more effectively, which raises more questions. Deeper inquiry into the effects of using remote assistance in social situations on PVI's social capital and success may lead to some interesting findings and have major implications for PVIs' social lives.

\section{Conclusion}
From our interviews with agents and users of Aira, we found four main contexts for which PVI use this service: object description, navigation, performance, and social interaction. Each of these comes with its own set of strategies and challenges. From these, we were able to deduce a set of characteristics that conversational agents should have in order to be successful in supporting PVI as well as design guidelines for these types of services. The overarching themes spoken about by agents and PVI alike are the respectful and equal collaboration between the two parties, mutual trust, and adaptability. These aspects and other characteristics of these conversational interactions turn out to have surprising nuances that contribute to the success of what can be difficult tasks by a team of two people who are not physically in the same space. 

Aira has proven to be quite effective, but there is potential for improvement in the exploration of parallel, nonverbal communication channels, like text and haptic cues. These additions may make the service usable in an even broader range of situations and increase the amount of visual information agents are able to deliver to PVI.

\bibliography{bibliography/sample}{}
\bibliographystyle{plain}

\end{document}